\documentclass[useAMS]{mn2e}
\usepackage[dvips]{graphicx}

\newcommand{\Msol}{$M_\odot$}

\newcommand{\lsim}{\raisebox{-3.8pt}{$\;\stackrel{\textstyle <}{\sim}\;$}}
\newcommand{\etal}{\mbox{{\rm et~al.\ }}}

\title[The black hole -- host galaxy relation in quasars]
{On the cosmological evolution of the black hole -- host galaxy 
relation in quasars}

\author[Portinari et al.]{Laura Portinari$^1$, 
Jari Kotilainen$^2$, Renato Falomo$^3$, Roberto Decarli$^{4,5}$\\
$^1$ Tuorla Observatory, Department of Physics and Astronomy, University 
of Turku, V\"ais\"al\"antie 20, FIN-21500 Piikki\"o, Finland\\
$^2$ Finnish Centre for Astronomy with ESO (FINCA), University 
of Turku, V\"ais\"al\"antie 20, FIN-21500 Piikki\"o, Finland\\
$^3$ INAF -- Osservatorio Astronomico di Padova, Vicolo dell'Osservatorio 5,
I-35122 Padova, Italy\\
$^4$ Universit\'a degli Studi dell'Insubria, Via Valleggio 11, I-22100 Como,
Italy\\
$^5$ Max Planck Institute for Astronomy, K\"onigstuhl 17, D-69117 Heidelberg,
Germany\\
E-mail: {\tt lporti,jarkot@utu.fi; renato.falomo@oapd.inaf.it; 
decarli@mpia-hd.mpg.de}
}
\begin{document}

\maketitle

\begin{abstract}
Quasars are useful tracers of the cosmological evolution of the black 
hole mass -- galaxy relation. 
We compare the expectations of Semi--Analytical Models
(SAM) of galaxy evolution, to the largest available datasets of quasar
host galaxies out to $z \simeq 3$. 

Observed quasar hosts are consistent with no evolution from the local 
$M_{BH} - L_{host}$ relation, and suggest a significant increase of the mass 
ratio $\Gamma = M_{BH}/M_\star(host)$ from $z=0$ to $z=3$. Taken at face value, 
this is totally at odds with the predictions of SAM,
where the intrinsic
$\Gamma$ shows little evolution and quasar host galaxies at high redshift
are systematically overluminous (and/or have undermassive BH).
However, since quasars preferentially trace very massive black holes 
($10^9-10^{10}$~\Msol) at the steep end of the luminosity and mass function, 
the ensuing selection biases can reconcile the present SAM with 
the observations. 
A proper interpretation of quasar host data thus requires the global approach 
of SAM so as to account for statistical biases.
\end{abstract}


\section{Introduction}

There is evidence that every galactic spheroid (elliptical galaxy or bulge) 
hosts a central 
supermassive black hole, with a strict relationship between the black
hole mass and the luminosity, mass, velocity dispersion, 
concentration and binding energy of the host 
(Kormendy \& Richstone 1995; Magorrian \etal 1998; Gebhardt \etal 2000;
Ferrarese \& Merritt 2000; Ferrarese 2002; Tremaine \etal 2002; 
Bettoni \etal 2003; H\"aring \& Rix 2004; Graham \& Driver 2007; 
Aller \& Rischstone 2007; Barway \& Kembhavi 2007). 
This discovery has highlighted the close connection between the process 
of galaxy formation at large, and the formation of the central black hole (BH),
endowed with its quasar activity; and is currently one of the major 
observational facts that the theory of galaxy evolution has to explain. 

In the Cold Dark Matter (CDM) hierarchical cosmological scenario, the usual 
paradigm is that (major) mergers are responsible for the joint origin and 
growth of black holes and galactic spheroids. Mergers trigger 
gas inflows feeding BH growth and quasar activity, while at the same 
time they modify the morphology of the galaxy into a bulge--dominated one
(e.g.\ Kauffmann \& Haehnelt 2000; Di Matteo \etal 2005). 
Alternative mechanisms link directly the BH growth 
to the intrinsic star formation activity or morphological evolution of the host
(e.g.\ Granato \etal 2001, 2004; Fontanot \etal 2006; Bower \etal 2006). All
these scenarios share an important feature: a quasar marks a very 
specific, short but crucial phase in the evolution of a galaxy. 
The host is expected to be a ``young spheroid'' 
where strong star formation (intrinsic or merger--induced) has just 
halted, by quasar feedback or by mere consumption of the cold gas 
that fed both the starburst and the quasar.
Thereafter the galaxy rapidly reddens and evolves passively,
while the central black hole becomes a ``dead quasar'' or a ``dormant black
hole'' (Springel \etal 2005a; Hopkins \etal 2008; Johansson \etal 2009ab)
--- until, possibly, later mergers or gas infall revive star formation 
and/or AGN activity.

On the observational side, major advances have been achieved in the past 
few years: a suitable number of detected quasar host galaxies at redshift 
{\mbox{$1<z<3$}} is nowadays available. Their luminosity apparently follows 
passive evolution, consistent with that of an elliptical galaxy formed at $z>3$
(Kotilainen \etal 2009), in contrast with the theoretical scenario
outlined above.
In this paper we aim at testing whether the predictions of current
merger--based models 
can be compatible with the available observations of quasar hosts.

Direct comparison to data on QSO host galaxies demands theoretical 
predictions on the properties of galaxies {\it specifically at the very phase 
of optical QSO activity}, as this is supposed to be a short but very 
critical phase of galaxy formation. The only explicit predictions in this sense,
in the framework of semi--analytical models, seem to date back to 
Kauffmann \& Haehnelt (2000); here we use the most recent public mock
catalogue from the Munich group to extract the expected properties of
quasar host galaxies, and compare them with the latest available data.

We consider in particular recent results on the evolution of the
BH mass --- host mass (or luminosity) relation. Peng \etal (2006)
and Decarli \etal (2010ab) find that the BH mass -- luminosity relation 
is roughly constant with redshift; considering the intrinsic fading of stellar
populations with age, this implies that the host stellar mass $M_\star$
associated to a given BH mass $M_{BH}$ decreases at high $z$. 
The evolution of the mass ratio $\Gamma=M_{BH}/M_\star$ is an important 
constraint on theoretical models, especially regarding the role of quasar 
feedback (Wyithe \& Loeb 2005; Fontanot \etal 2006).

The outline of the paper is as follows. In Section~2 we describe 
the semi--analytical models in use and how quasar host galaxies are
selected from the mock galaxy catalogues. In Section~3 and~4 we discuss the
evolution of the BH mass--stellar mass and of the BH mass--luminosity relation,
compared to observational evidence. In Section~5 we discuss the 
mass function of BH in quasars at high redshift.
In Section~6 we outline our conclusions and suggestions for future studies. 
In the Appendix we discuss the problem of transforming observed host 
luminosities into stellar masses, and the significance of their apparent
passive evolution.
\section{Merger--triggered quasar activity: 
Semi--Analytical Models}

For about a decade semi--analytical models (SAM), superposing the evolution 
of visible structures over that of the underlying CDM, treated galaxy 
formation (White  \& Rees 1978) and quasar activity (Efstathiou \& Rees 1988) 
separately.
After growing evidence of the black hole--host bulge relation, the two
lines of investigation merged: galaxy evolution models 
have incorporated BH growth and AGN activity. The first ``unified'' model
was by Kauffmann \& Haehnelt (2000), followed by many others
(Enoki \etal 2003; Granato \etal 2004; Cattaneo \etal 2005; Menci \etal 2006; 
Croton \etal 2006; Bower \etal 2006; Fontanot \etal 2006; Malbon \etal 2007;
Somerville \etal 2008; Marulli \etal 2008; Bonoli \etal 2009;
Jahnke \& Macci\'o 2010; Fanidakis \etal 2011).

Most of these models assume that the joint origin of spheroids and black
holes is a consequence of mergers. In few cases, central BH accretion 
is (also) associated to the intrinsic evolution of the host:
to its star formation activity (Granato \etal 2004; Fontanot \etal 2006)
or to its morphological transformation from disc to bulge (Bower \etal 2006;
Fanidakis \etal 2011). 
Another important distinction among the various models is whether quasar 
feedback at high redshift plays a key role  
(e.g.\ Granato \etal 2004; Fontanot \etal\ 2006; Menci \etal 2006; 
Somerville \etal 2008) or not.\footnote{Attention has recently focussed 
on the role of AGNs in halting cooling flows in massive galaxies and clusters
at low redshift, to better reproduce their red colours 
and the bright end of the local luminosity function: the ``radio mode'', 
associated with low--level accretion (Croton \etal 2006; 
Bower \etal 2006, 2008; Kawata \& Gibson 2005). Here we refer to feedback
in the ``quasar mode'', related to the bright phase of quasar activity at high 
redshift, where the bulk of BH growth and quasar energy emission 
occurs. Notice that effective quasar feedback is directly supported 
by recent observations of outflows of molecular gas
(Feruglio \etal 2010; Sturm \etal 2011).}

Our discussion relies on the public catalogue 
of SAM galaxies by the Munich group (De Lucia \& Blaizot 2007), based on the 
Millennium simulation (Springel \etal 2005b) and retrievable from the 
Millennium database\footnote{http://www.g-vo.org/Millennium}. 
As to the ``quasar mode'' BH accretion at high redshift, this SAM
follows essentially the recipe of its prototype Kauffmann \& Haehnelt 
(2000; see also Croton \etal 2006). 
Each merger triggers a starburst, and a few percent of the available 
cold gas mass $m_{cold}$  accretes onto the central BH:
\begin{equation}
\label{eq:DMBH}
\Delta M_{BH}=f_{BH} \frac{m_{sat}}{m_{cen}} 
\frac{m_{cold}}{1+(280~{\rm km~sec^{-1}/V_{vir}})}
\end{equation}
The mass of the resulting BH is the sum of the progenitor BH masses, and
of the (dominant) accreted mass $\Delta M_{BH}$.
The parameter $f_{BH}=0.03$ is tuned to reproduce the observed local 
BH mass--bulge mass relation at $z=0$. The efficiency of BH growth scales
with the mass ratio  $m_{sat}/m_{cen}$ of the merging galaxies (``satellite''
and ``central'') so that the fractional contribution of minor mergers 
to quasar activity is small. BH accretion in the quasar mode is thus dominated 
by major mergers (mass ratio larger than 1:3) which result in the formation 
of a spheroid.

QSO activity in this model is always associated to a recent merger 
and active star formation.
Quasar activity is a by-product of the merger, with no impact on the evolution 
of the galaxy --- arguing that any quasar--induced feedback can be formally 
included in the strong supernova feedback 
accompanying the starburst. The Munich SAM effectively belong to the 
no-feedback category in the quasar mode.

The Munich SAM series has been successfully tested and tuned to reproduce
a wide range of observational properties of the galaxy population, such as:
galaxy clustering (Springel \etal 2005b); galaxy luminosity function, 
colour and morphology distributions, colour--magnitude, mass--metallicity 
and Tully--Fisher relations, cosmic star formation and BH growth history 
(Croton \etal 2006); the formation history of elliptical galaxies
(De Lucia et al.\ 2006); the properties of Bright 
Cluster Galaxies (De Lucia \& Blaizot 2007). This SAM is optimized to describe 
the galaxy population, but results on the corresponding AGN population are 
discussed by Marulli \etal (2008) and Bonoli \etal (2009).
The cosmological evolution of the $M_{BH}-M_{bulge}$ relation in this model 
is discussed by Croton (2006).

In our study we use the available public mock galaxy catalogue 
of the Munich SAM (De Lucia \& Blaizot 2007) to discuss the evolution 
of the scaling relations (BH mass versus host mass and luminosity) 
{\it as traced specifically by quasar host galaxies} up to $z=3$.

\subsection{Quasar host galaxies in the Munich SAM}
\label{sect:selection}

To compare the Munich SAM to observed data on the BH--host relation in
quasars, we need to know, at each redshift/snapshot of the SAM: 
(a) which are the active galaxies, (b) their BH masses and (c) their stellar 
masses and luminosities. All of this information is directly available 
in the public catalogue of De Lucia \& Blaizot (2007), with no need for further
assumptions. 
The active galaxies in ``quasar mode'' are those that have just suffered a 
merger; we query the database to select recent mergers following the
example instructions provided on the web-site. 
``Recent merger'' in this case means, merged
since the previous redshift snapshot, typically 
$1-3 \times 10^8$~yrs before. This is longer than the duty cycle 
of optical quasar activity ($10^7-10^8$~yrs) so that we can identify 
the very moment of quasar shining  only approximately--- but it's as close 
as we can get with the time resolution available in the public SAM catalogue.

For the recent merger/quasar mode galaxies we retrieve 
the following information: BH mass, stellar mass, gas mass and luminosity 
in various bands. We also retrieve the BH, stellar and gas mass of the
progenitors: this gives us the BH mass growth $\Delta(M_{BH})$ (from the 
mass difference between the progenitor BHs and the resulting BH) and 
the merger mass ratio. We also retrieve BH and galaxy properties for the
overall galaxy population, to discuss differences (in luminosity mainly,
see Section~\ref{sect:hostlum}) with the quasar host subset.

The quasar population and AGN luminosity function associated with these same
quasar hosts, was studied by Marulli \etal (2008) by adding to the SAM
various prescriptions about the quasar light curve associated to $M_{BH}$ and 
$\Delta(M_{BH})$ in each merger. Notice that their (or any) additional 
assumptions on the quasar light curve do not affect the basic 
quantities (BH masses, $\Delta(M_{BH})$, galaxy properties etc.) available 
in the public mock catalogue, that set the scaling relations in the SAM. 
We comment later on the results of Marulli et al.\ (2008) in relation to ours, 
but shall not develop here a new model for the quasar population and light
curves as it is not needed to study the scaling relations.

For practical reasons (avoid overload of unneccessary output data
from the database query) we impose some additional restrictions
that do not affect the substance of the quasar host population.
\begin{enumerate} 
\item
We consider mergers with a mass ratio (in cold baryons, i.e.\ stellar mass
 + cold gas mass) larger than $1:9$. As major mergers 1:3 largely dominate 
BH growth (Croton et al. 2006), 1:9 is a very safe limit to include all 
significant optical QSO activity --- considering that the latter does 
correspond to the bulk of the BH growth (Soltan 1982; Yu \& Tremaine 2002).
We checked that, among our final selected objects, major mergers 
(with mass ratio larger than 1:3) contribute about half of the quasar hosts 
with $M_{BH} =10^8$~\Msol\ and dominate by 70--80\% at the massive end, 
$M_{BH} \geq 10^9$~\Msol. The quoted percentages are stable with redshift.
\item
We restrict to galaxies hosting a BH mass 
$M_{BH} \geq 2 \times 10^7$~\Msol; this a conservative choice that fully
covers the BH mass range of the observational dataset (QSO hosts 
at high $z$ have $M_{BH} \geq 10^8$~\Msol) even including the 0.4~dex error
on the measured $M_{BH}$, discussed later in Section~4. Besides, 
BH masses below our adopted limit hardly contribute to the optical
quasar population (see e.g.\ McLure \& Dunlop 2004; Shankar \etal 2010); 
for instance, in the latest SDSS quasar sample of Shen \etal (2011),
only 19 out of over 22.000 BH masses measured with H$_\beta$ lines
are below $2 \times 10^7$~\Msol.\\
Considering specifically the observational QSO sample of Decarli \etal 
(2010a), all objects at $z>0.5$ have $M_V<-24$, which is much brighter
than expected from our adopted mass cut. Indeed a BH of 
$2 \times 10^7$~\Msol, emitting typically around 0.5 of its Eddington 
luminosity (McLure \& Dunlop 2004; Labita \etal 2009) shines with 
$L_{bol}=1.3 \times 10^{38}$~W, corresponding to $M_B=-22.02$ (McLure \&
Dunlop 2004) or $M_V=-22.24$ (assuming a typical quasar colour $B-V$=0.22,
from Cristiani \& Vio 1990). Clearly our mass cut covers both the mass 
and luminosity range relevant for comparison to observations.
\item
We neglect multiple mergers of three or more progenitors, for simplicity 
in the treatment of the query output (multiple mergers appear as a repeated 
double merger in the output list).
We also neglect mergers with progenitors identified too early on
(two or more snapshots before, rather than in the immediately previous 
snapshot) as the instant of the merger and the corresponding quasar activity 
is not guaranteed to be very recent, i.e.\ the time resolution on the
quasar host phase is much worse. These two criteria together exclude 
less than 10\% of the merger events, bearing no impact on our discussion.
\item
As the Soltan argument indicates that optical QSO activity traces
the bulk of the BH growth, we test the additional requirement that the
selected mergers induce a BH growth of more than 50\% --- a simple, reasonable 
way to ensure that the merger corresponds to significant quasar activity.
We verified that most of our conclusions are not affected when relaxing 
this ``doubling'' criterion; when this is the case, both alternatives are shown 
(Section~5).
\end{enumerate}
The selected mergers/quasar hosts represent 5-6\% of the global galaxy 
population at $z \geq 1$, and 2\% at $z=0.5$. At each redshift snapshot 
between $z=1$ and 3, our discussion is based on a sample of $1-3 \times 10^4$ 
merger galaxies selected as above, out of a global galaxy population 
of $3-5 \times 10^5$ objects.

Beyond $z \sim 1$ mergers are usually considered the main trigger of AGN
activity, while at lower redshift other mechanisms are likely to contribute
or even dominate (secular evolution and bar--driven instabilities; mass loss 
from old stellar populations; e.g.\ Hopkins \& Hernquist 2009; 
Kauffmann \& Heckman 2009; Cisternas \etal 2010). Therefore, our selection 
of recent mergers (and the underlying assumptions in the SAM about 
quasar activity) may be not well suited for AGN hosts at $z<1$; 
but in this paper
we are mostly concerned with the hosts of bright quasars at high redshift. 

\begin{figure*}
\includegraphics[width=0.9 \textwidth]{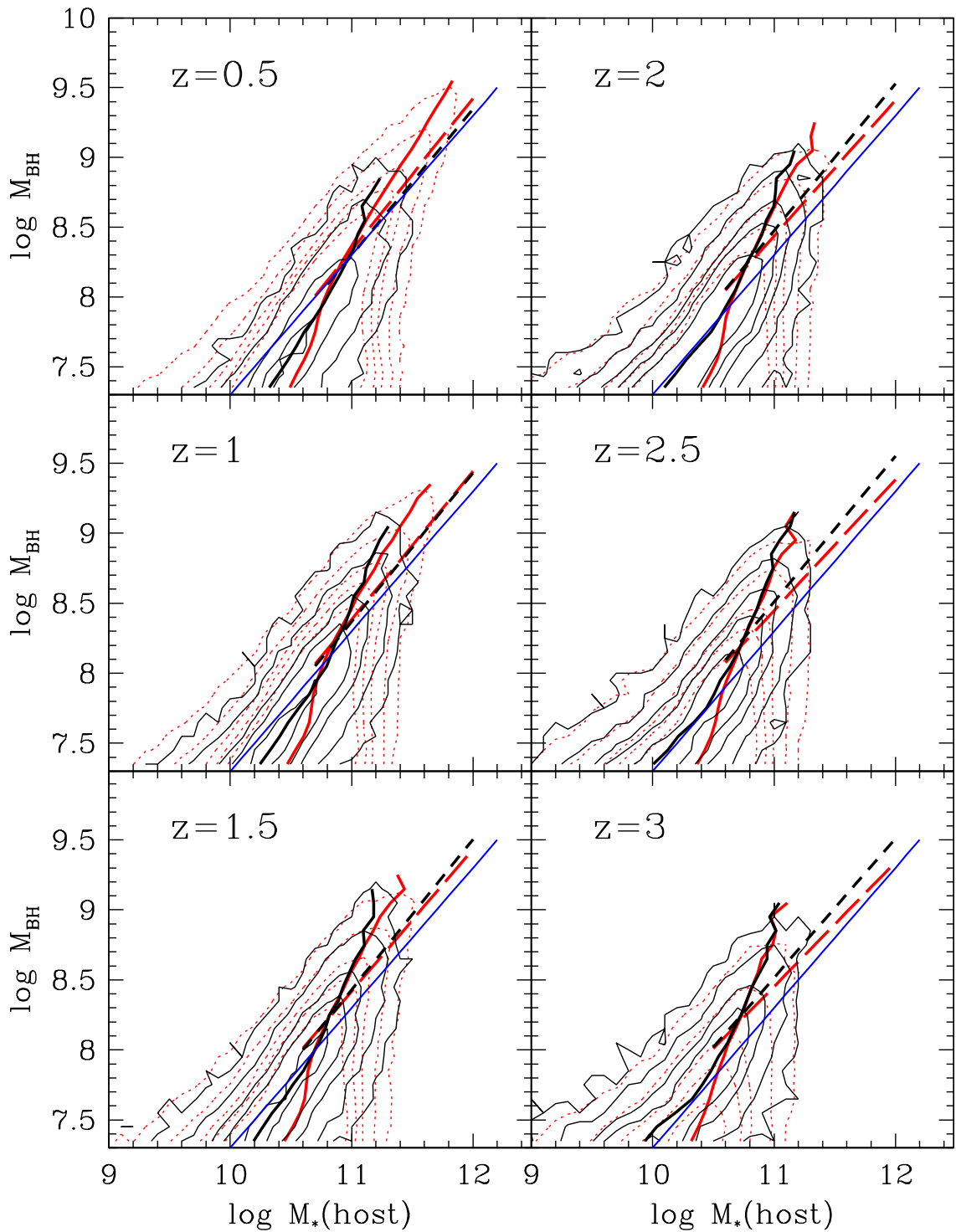}
\caption{Relation between BH mass and host stellar mass at various redshifts, 
as derived from the SAM galaxy catalogue of De Lucia \& Blaizot (2007) 
in the Millennium database.
{\it Dotted (red) contours}: isodensity contour plots for the global galaxy 
population.
{\it Solid contours}: selected quasar hosts (recent mergers with significant 
BH accretion, see text). The contour levels for the far more numerous 
global galaxy population are 10 times those of the quasar hosts.
The solid lines trace the median host luminosity as a function of BH mass, 
for the global galaxy population and for the quasar hosts.
The dashed lines trace the bisector fit relations: long--dashed (red) line 
for the global galaxy population, short--dashed for quasar hosts; 
both are defined for $M_{BH} \geq 10^8$~\Msol (the minimum BH mass 
relevant for comparison to observed high $z$ QSO hosts) but this limit is not 
crucial for the resulting relation. The (blue) thin straight line 
is the observed 
relation at $z=0$: $M_{BH}/M_\star(bulge)$=0.002 (Marconi \& Hunt 2003).}
\label{fig:MBH_Mstar_all}
\end{figure*}

Furthermore, at high redshift it is observationally hard to decompose
the host galaxy into its bulge/disc component, so the observed scaling relations
often refer to the global host galaxy (a recent exception is Bennert \etal
2011). For consistency with this limitation, we extract from the SAM the 
scaling relations between BH and host galaxy, rather than host spheroid. 
However, as customary in the observational papers, we shall compare the high 
redshift results for the host {\it galaxies} with the $z=0$ relation between 
BH and host {\it bulge} (Marconi \& Hunt 2003; H\"aring \& Rix 2004).

\section{The BH mass--host mass relation}
\label{sect:hostmass}

In this section we discuss SAM predictions on the evolution of the 
BH mass--host mass relation. Fig.~\ref{fig:MBH_Mstar_all} shows the
distribution, in the $M_{BH} - M_\star(host)$ plane, of quasar hosts
(solid contours) and of the global galaxy population (dotted contours)
at various redshifts. In this plane, the two populations occupy the same
loci, i.e.\ QSO hosts are a fair sample of the general galaxy population
(at least for $M_{BH} \geq 10^8$~\Msol, the relevant range for high--$z$ 
observed quasars).

To discuss the evolution of the $M_{BH} - M_\star(host)$ relation,
we need to specify how the relation can be defined in the models.
From the physical point of view, neither BH mass nor host stellar mass
can be selected to be the independent versus dependent variable, 
as they both are the result of a third process: galaxy formation and evolution.
For this sort of related variables, the best statistical tracer of the 
intrinsic mutual relation is a bisector fit relation (Isobe \etal 1990;
Akritas \& Bershady 1996). This definition is also the one
adopted for the observed relation in the local Universe (Marconi \& Hunt 2003; 
H\"aring \& Rix 2004).
The ``intrinsic'' (bisector fit) relation for the SAM galaxy catalogue 
(dashed lines in Fig.~\ref{fig:MBH_Mstar_all}) at low redshift matches very 
well the local relation observed at $z=0$; and displays little
evolution with redshift. The latter is a general feature 
of SAM that do not include quasar feedback 
(Wyithe \& Loeb 2005; Fontanot \etal 2006; Malbon \etal 2007).

Notice that the slope of the bisector fit relation in the 
$\log (M_{BH}) - log (M_\star)$ plane turns out to be always close to 1;
therefore, in practice this definition is very similar to
what we would obtain with the more common approach of fixing the slope to 1 and 
fitting a unique value for the ratio $\Gamma=M_{BH}/M_\star$ (e.g.\ Croton 2006;
Decarli \etal 2010). We also notice that, for the same SAM models considered
here, Croton (2006) report a significant evolution in the $M_{BH}-M_{bulge}$
relation; this is not in contrast with our findings: most of the evolution 
he reports is due to the redistribution of stars from the disc to the bulge 
component, an effect which largely cancels out when we consider the global 
host galaxy.

\begin{figure}
\includegraphics[angle=-90,width=8.8truecm]{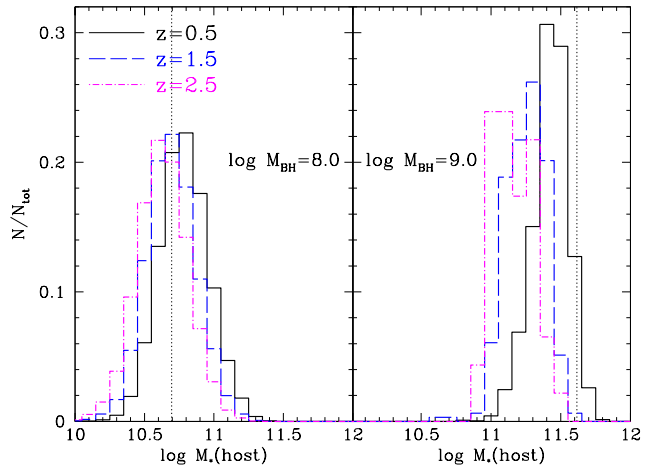}
\caption{Histograms of the distribution of host galaxy masses corresponding 
to a given BH mass, as a function of redshift. The dotted vertical lines mark 
the host mass predicted by the intrinsic bisector-fit relation 
(at $z=0.5$, but evolution with redshift is negligible). The offset between
the histograms and the vertical line represents the Lauer bias. The plot refers
to the global galaxy population in the SAM catalogue; QSO hosts behave in a
very similar way.}
\label{fig:lauer}
\end{figure}

The negligible evolution of the intrinsic (bisector fit) relation appears
in contrast to observational results, when taken at face value (e.g.\
Peng 2006; Decarli \etal 2010ab).
When comparing to high redshift data, however, we must take into account
that quasar hosts are operatively detected starting from QSO selected samples,
and tend to pick out the median host mass 
as a function of BH mass (solid lines in Fig.~\ref{fig:MBH_Mstar_all}). 
The latter definition of the $M_{BH}-M_\star(host)$ relation mimics more 
closely the empirical sampling, and also traces better the contour plots, 
which are a convolution between the intrinsic BH mass--host mass relation, 
its scatter, and the mass function of galaxies (Lauer \etal 2007). 
There is a systematic bias between the two definitions of the relation: 
the more luminous quasars tend to trace 
over--massive BH with respect to the underlying intrinsic BH--host relation. 
This is due to the fact that, being massive galaxies very rare, the most 
massive BH are more easily found as outliers hosted in undermassive 
(but more frequent) hosts. 
This bias is discussed 
extensively by Lauer \etal (2007) and we shall refer to it as the Lauer bias. 
The bias can be defined either as an excess of BH mass at a given host 
mass/luminosity/velocity dispersion; or as an offset in host properties 
at given BH mass. To interpret quasar host data, where the effective 
independent variable in the selection is the BH mass of the QSO, we prefer
the latter approach: $\Delta \log M_\star$ is the offset in host mass between 
the median 
relation marginalized over BH mass, and the intrinsic (bisector fit) relation. 
The Lauer bias for the global galaxy population in the SAM catalogue 
is represented in Fig.~\ref{fig:lauer} and Table~\ref{tab:lauer}.
In these SAM, the deviation of the distribution from the intrinsic relation is
significant ($\geq$0.2~dex in $M_\star(host)$, i.e.\ larger 
than the typical dispersion) around $M_{BH}=10^9$~\Msol. At this BH mass, 
the bias increases from 0.2~dex to 0.6~dex between $z=0.5-3$; this is
comparable to the evolution determined by Decarli \etal (2010b), considering 
that most of their objects at $z>1$ indeed have $M_{BH} \geq 10^9$~\Msol. 

\begin{table}
\caption{Lauer bias for the global galaxy population in the SAM catalogue.
We indicate the offset $\Delta \log M_\star(host)$ (typically an underestimate: 
minus sign) of the median host stellar mass at a given BH mass, with respect 
to the intrinsic bisector fit relation. 
The dispersion is estimated from the 16 and 84 percentiles of the distribution
(corresponding to 1 standard deviation for a gaussian distribution).
For the entry in the bottom right corner ($z=3$, $M_{BH}=10^9$~\Msol), due to the
small number of objects we considered the average 
logarithmic host mass and the extreme values in the sample.} 
\begin{tabular}{l l l l}
\multicolumn{1}{c}{$z$} & $\log M_{BH}=8$ & $\log M_{BH}=8.5$ & $\log M_{BH}=9$ \\
\hline
0.5 & $0.08 \pm 0.18$ & $-0.08 \pm 0.15$ & $-0.19 \pm 0.13$ \\
1.0 & $0.09 \pm 0.18$ & $-0.09 \pm 0.14$ & $-0.23 \pm 0.15$ \\
1.5 & $0.10 \pm 0.18$ & $-0.12 \pm 0.15$ & $-0.32 \pm 0.15$ \\
2.0 & $0.11 \pm 0.18$ & $-0.15 \pm 0.15$ & $-0.37 \pm 0.17$ \\
2.5 & $0.11 \pm 0.18$ & $-0.17 \pm 0.15$ & $-0.45 \pm 0.16$ \\
3.0 & $0.11 \pm 0.18$ & $-0.21 \pm 0.16$ & $-0.58 \pm 0.09$ \\
\hline
\end{tabular}
\label{tab:lauer}
\end{table}

\subsection{The evolution of $\Gamma$}
\label{sect:Gamma}

\begin{figure}
\includegraphics[angle=-90, width=8.8truecm]{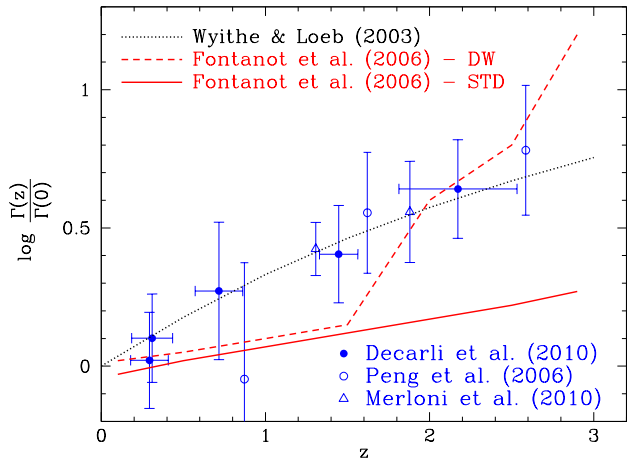}
\caption{Symbols with errorbars represent the evolution of the mass ratio
$\Gamma=M_{BH}/M_\star$ 
in quasar host galaxies, from various observational papers, as compiled 
by Decarli \etal (2010b). Lines represent the predictions of various SAM from
the literature. Wyithe \& Loeb (2003) and the Drying Wind model of Fontanot 
\etal (2006) include self-regulation by quasar feedback, while the 
STandarD model of Fontanot \etal does not.}
\label{fig:Gamma_models}
\end{figure}

The cosmological evolution of the BH/host mass ratio:
\[ \Gamma = \frac{M_{BH}}{M_\star(host)} \]
can contribute to discriminate between different scenarios of 
co--evolution of central super--massive black hole and host galaxy: 
models where quasar feedback plays a prominent role predict a stronger 
evolution in $\Gamma$ (increasing in the past) than models
that do not include this effect (Wyithe \& Loeb 2005); and different feedback
scenarios result in different predictions for $\Gamma(z)$ (Fontanot 
\etal 2006). It is thus tempting to conclude that the strong evolution 
detected in recent observational studies favours the models that take
feedback and self--regulation into account (Fig.~\ref{fig:Gamma_models}). 
In particular, it should exclude ``extreme 
merger scenarios'' where the relation between BH mass and host mass is just 
the statistical outcome of the stochastic merger history, with no 
direct physical relation between black hole and bulge formation at the level 
of individual galaxies (Peng 2007; Jahnke \& Macci\`o 2010).

However, an apparent evolution of $\Gamma$ is seen in the SAM due to
the Lauer bias, as the combination of two factors:
(i) the slope of the median $M_{BH}-M_\star(host)$ relation is steeper than 1:1
(closer to 2:1)
and (ii) the mass function of quasars and the Malmquist bias affect 
the accessible parameter range one can address as a function of redshift.
We sample more luminous and massive quasars at increasing redshift and 
tendentially find smaller hosts and larger $\Gamma$.

\begin{figure}
\includegraphics[angle=-90,width=8.8truecm]{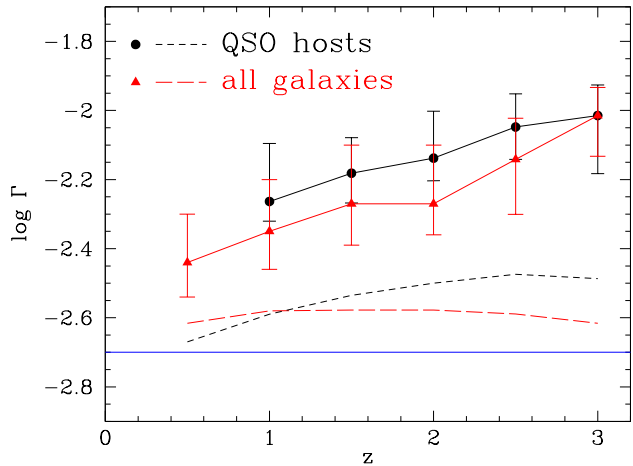}
\caption{Evolution with redshift of $\Gamma$, for QSO hosts and for all 
galaxies in the SAM catalogue. The dashed lines refer to the bisector fit 
relation. Symbols connected with solid lines refer to the median $\Gamma$
for objects with a BH mass around $M_{BH}=10^9$~\Msol; the errorbars indicate 
the 16 and 84 percentiles of the distribution. The horizontal  
line marks the local value $\Gamma=0.002$ (Marconi \& Hunt 2003).}
\label{fig:Gamma_evol}
\end{figure}

Fig.~\ref{fig:Gamma_evol} illustrates that, when derived from the intrinsic 
relation (bisector fit, dashed lines), $\Gamma$ is close to the local reference
value with little evolution (about 0.2~dex offset between $z=0$ 
and $z=2-3$). 
In contrast, the median $\Gamma$ at $M_{BH}=10^9$~\Msol\ shows
a significant offset (a factor of 2--3 already at low redshift) and evolution 
with respect to the local value. This apparent evolution of $\Gamma$ due 
to the Lauer bias is comparable to that traced by the data in 
Fig.~\ref{fig:Gamma_models}, considering that observational samples mostly
include QSOs with $M_{BH} \geq 10^9$~\Msol. 
This suggests that the $\Gamma$ evolution inferred from 
the observations may be largely due to the bias, and be compatible
even with models that do not include effective quasar feedback.

Decarli \etal (2010b) performed a more empirically--based estimate 
of the Lauer bias expected in their data and found it to be negligible 
with respect to the observed evolution.
The extent of the Lauer bias depends on the luminosity/mass function 
of galaxies and of super-massive black holes, on the scatter of the intrinsic 
relation and on its evolution with redshift (Lauer \etal 2007). 
For the SAM considered here, there is evidence 
(see Section\ref{sect:QSOmassfunc}) that the models underestimate the number 
of massive quasars 
at high $z$; consequently, the Lauer bias in the SAM is probably 
exhacerbated and ``shifted'' at proportionally too low BH masses.
Nontheless, our results show that it is an important ingredient in the
interpretation of the data, and the global approach provided by SAM 
is needed to interpret the properties of quasar host samples.

\begin{figure*}
\includegraphics[width=0.9 \textwidth]{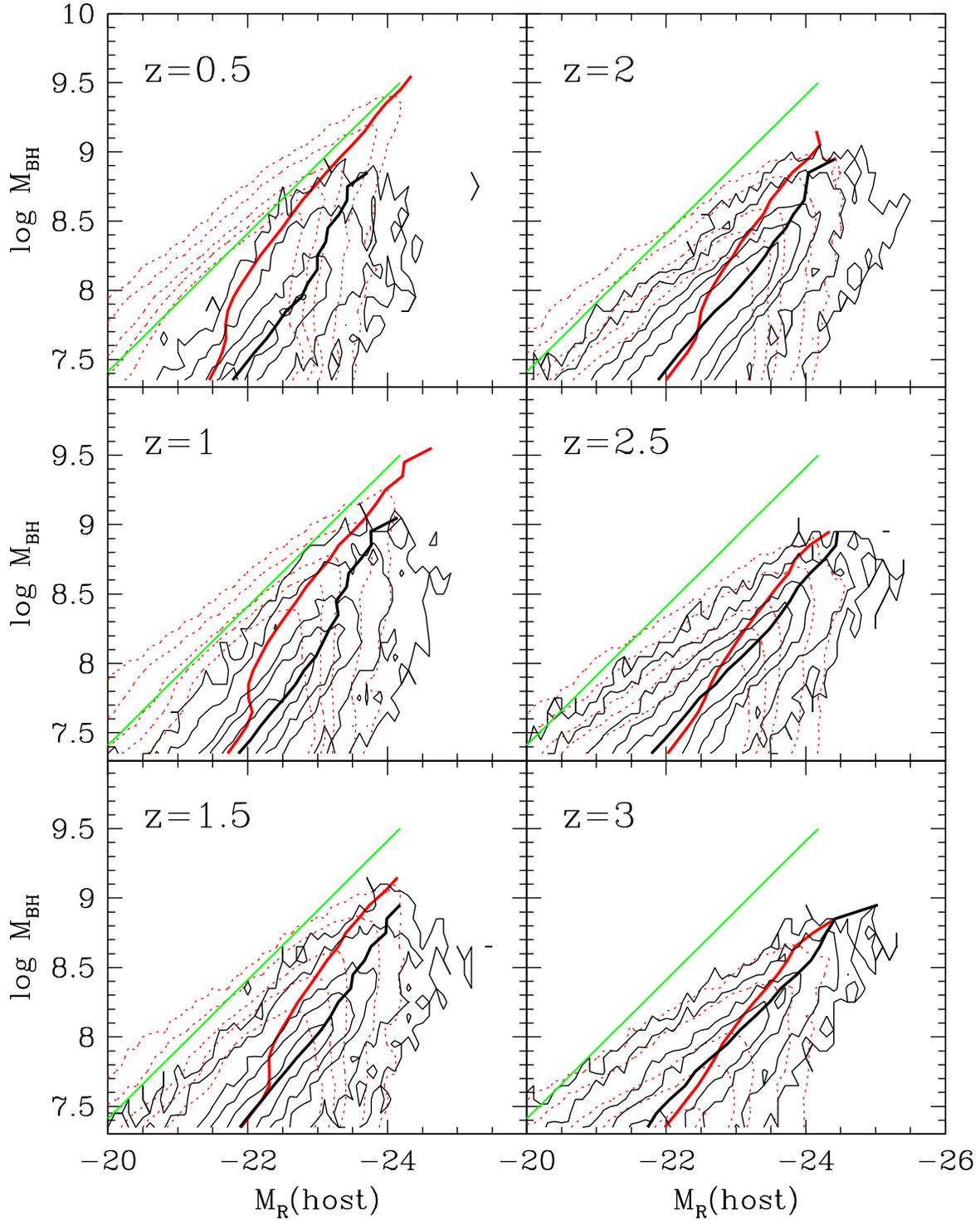}
\caption{Evolution with redshift of the relation between BH mass and host 
R--band magnitude (including dust extinction) in the SAM galaxy catalogue.
As in Fig.~\protect{\ref{fig:MBH_Mstar_all}}, the (red) dotted and the solid
contours refer to the global galaxy population and to the QSO hosts, 
respectively; the solid lines show the corresponding median
relations. The (green) thin straight line is the observed relation at $z=0$ 
(Bettoni \etal 2003, adapted to the cosmology of the Millennium run 
with $h=0.73$).}
\label{fig:MBH_MRhost_all}
\end{figure*}

\begin{figure*}
\includegraphics[width=0.8 \textwidth]{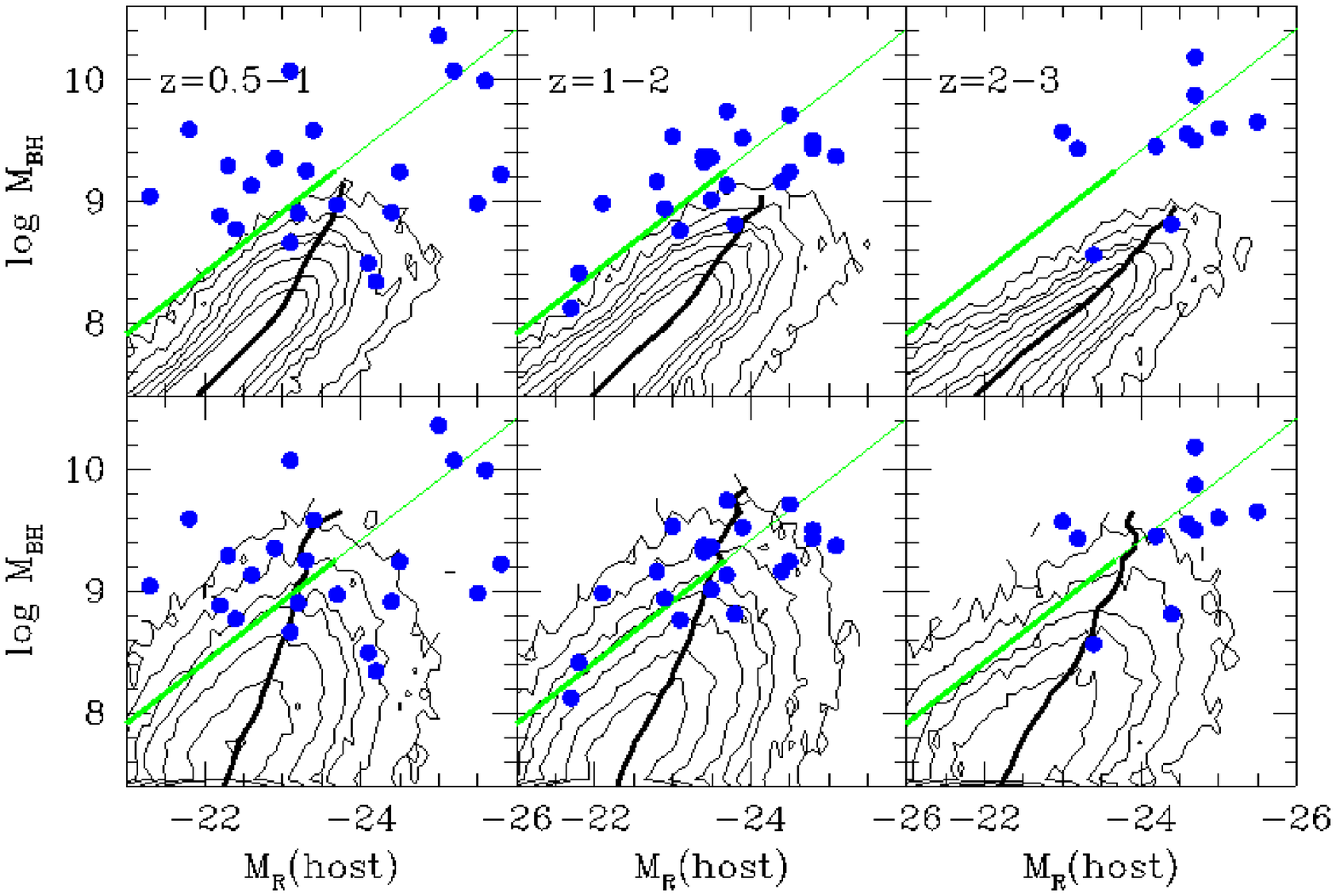}
\caption{Evolution with redshift of the relation between BH mass and host 
luminosity in SAM quasar hosts, compared to 
observations (Decarli \etal 2010ab; dots) in three redshift bins. 
The light (green) straight line is the local relation at $z=0$ (Bettoni 
\etal 2003),
extended with a thin line at magnitudes brighter than $M_R=-24$.
{\bf Top panels:} actual SAM quasar hosts; {\bf bottom panels:} convolving 
SAM predictions with observational errors (1$\sigma$) of 0.3~mag in M$_R$(host)
and 0.4~dex in log(M$_{BH}$).}
\label{fig:MBH_MRhost_obs}
\end{figure*}

\section{The BH mass--host luminosity relation}
\label{sect:hostlum}

The BH mass---host mass relation is physically more
meaningful, yet the most direct comparison between models and data is for 
the BH mass --- host luminosity relation. 
Observationally, in fact, we measure the luminosity 
of detected quasar host galaxies.
Their stellar mass is then derived indirectly, typically assuming that
the host is a spheroidal galaxy evolving passively since a higher formation 
redshift (Peng \etal 2006; Kotilainen \etal 2009; Decarli \etal 2010b). 
This is a quite different picture from the ``young spheroid'' scenario
of theoretical models. Further differences in the adopted stellar Initial Mass 
Function (IMF) can easily introduce systematic offsets up to 0.3~dex 
in the $M_\star/L$ ratio (Bell \& de Jong 2001; Portinari \etal 2004).
The issue is further discussed in the Appendix.

Therefore, in this Section we compare directly SAM to observational data
in the BH mass --- host luminosity plane. 
We consider the rest--frame $R$ band magnitude which is the most common band 
of choice in the observational datasets.\footnote{For 
comparison to observational data, we have transformed the Johnson $R$--band 
magnitudes provided for the SAM in the Millennium database, to Cousins $R$-band
magnitudes. We have used $(V-R)_C=0.715 (V-R)_J -0.02$ (Bessel 1983), valid
up to $(V-R)_C=0.8$ which fully covers the colour range spanned by the SAM 
galaxies. 
Galaxies are ``fainter'' in Cousins $R$ band and bluer 
in $(B-R)_C$, $(V-R)_C$ colours; the filter corrections range between 0.1~mag
for the bluest objects (QSO hosts at high redshíft, with typical 
$(V-R)_C \geq 0.2$) and 0.25~mag for the reddest ones (general galaxy 
population at $z=0$, with typical $(V-R)_C < 0.55$).} 
In Fig.~\ref{fig:MBH_MRhost_all} we show the locus of SAM galaxies in the 
R band magnitude --- BH mass plane, at various redshifts. In this plane,
quasar host galaxies are \underline{not} a fair sample 
of the global galaxy populations: having suffered a recent merger with 
associated starburst, they tend to be overluminous and bluer than average.
Indeed at low redshifts, quasar hosts in the SAM are systematically brighter 
by about 0.5~mag, at a given BH mass.
At higher redshifts however ($z > 2$), due to the younger age and more intense 
star formation activity of the galaxy population at large, the offset between 
the two populations tends to vanish. 

In Fig.~\ref{fig:MBH_MRhost_all}, we see that at low $z$ the median relation
for the global galaxy population (thick solid line tracing the dotted contours)
agrees with the relation
observed in the local Universe (this straight line), while departing from it 
at higher redshift. Quasar hosts are always overluminous than the local 
relation, at any redshift.
Both trends appear to be at odds with observations, that indicate a 
non--evolving
BH mass---luminosity relation (Peng \etal 2006; Decarli \etal 2010b). 

\begin{figure*}
\includegraphics[angle=-90,width=8.8truecm]{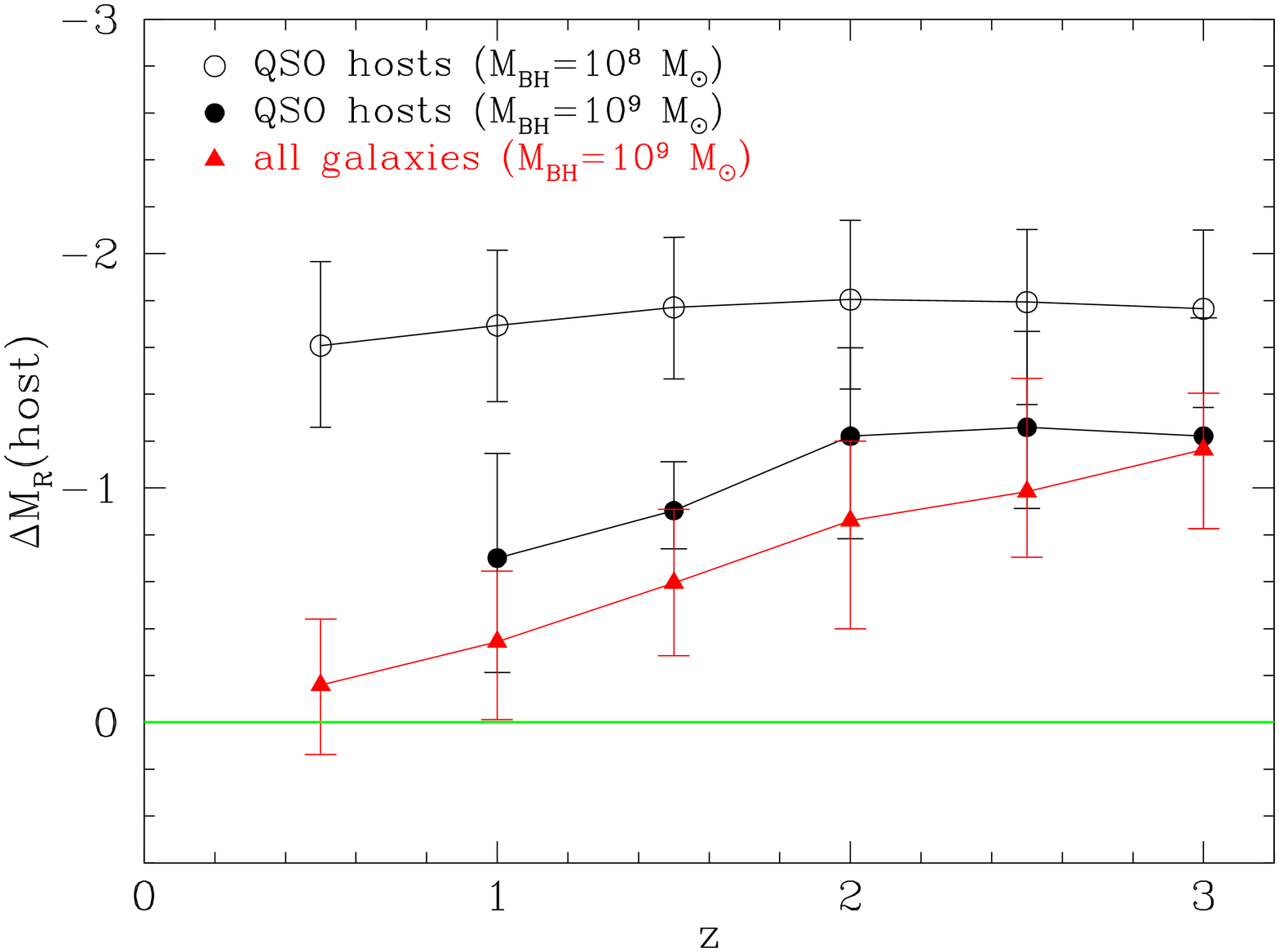}
\includegraphics[angle=-90,width=8.8truecm]{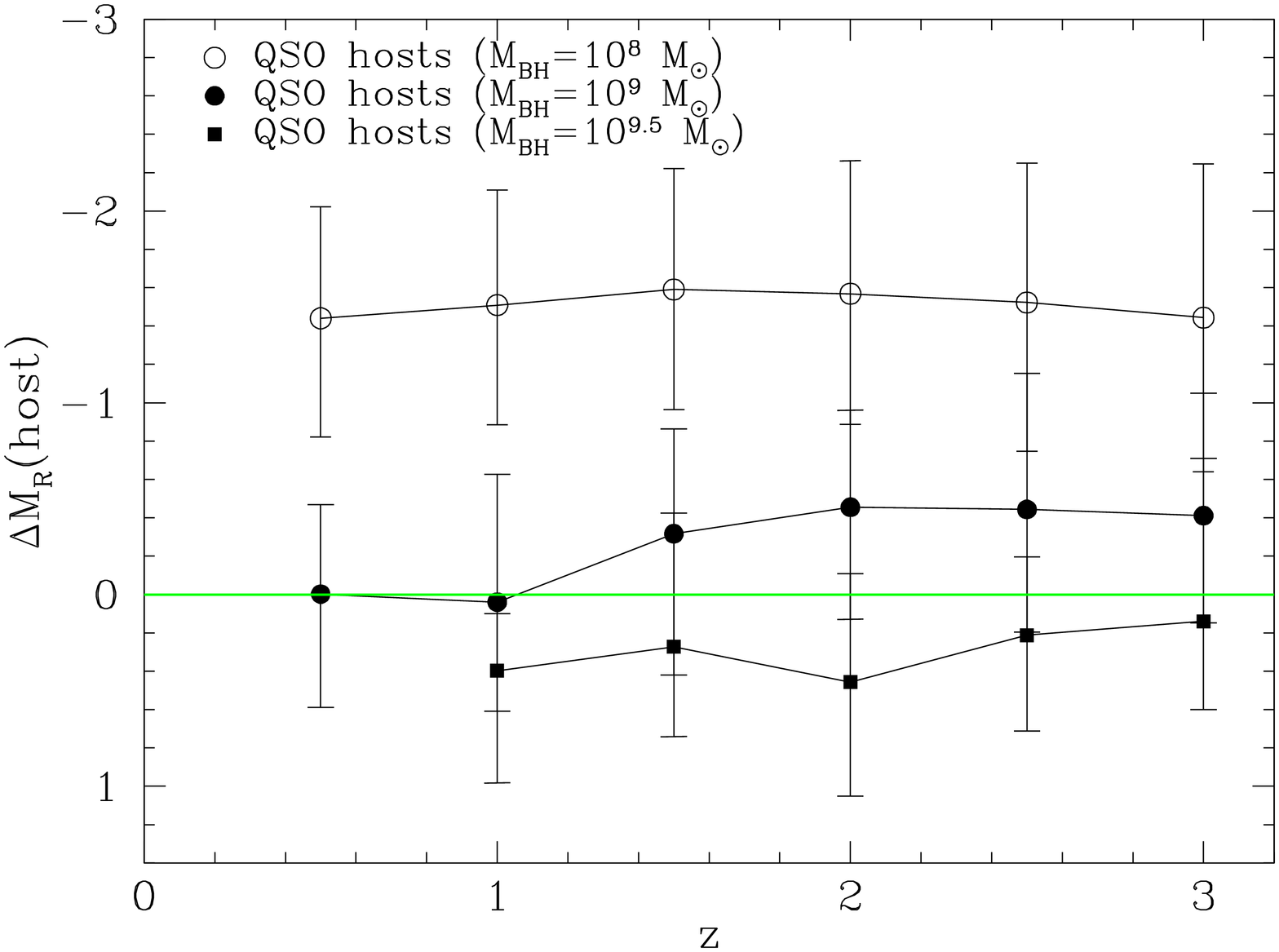}
\caption{SAM predictions on the evolution of the offset $\Delta M_R$ 
with respect to the local BH mass -- host luminosity relation (corresponding
to luminosity $M_R=-21.2$,--23.2,--24.2 for $M_{BH}=10^8$, $10^9$, 
$10^{9.5}$~\Msol; see Fig.~\protect{\ref{fig:MBH_MRhost_all}}). 
The errorbars indicate the 16 and 84 percentiles of the distribution.
{\bf Left panel}: ``real'' evolution in the SAM, for QSO hosts and for the 
general galaxy population; {\bf right panel}: including convolution 
with observational errors for QSO hosts.}
\label{fig:DMR_evol}
\end{figure*}

This discrepancy is evident in Fig.~\ref{fig:MBH_MRhost_obs} (top panels) where
we compare directly the observations of Decarli \etal (2010ab) to the properties
of SAM quasar hosts in the corresponding redshifts range. 
At given BH mass, the model QSO hosts are clearly overluminous with respect 
to the data; and/or SAM produce undermassive BH at given host luminosity. 
Even considering
that the normalization of the measured BH masses is somewhat arbitrary, 
depending on the assumed geometry of the broad line regions, one can hardly
reconcile model predictions with the data: the minimum BH masses, corresponding
to the isotropic case, would be systematically lower by 0.5~dex than the
normalization adopted by Decarli \etal (2010a); but a disc--like geometry
is favoured by a number of arguments (Decarli \etal 2008ab; 
Graham et~al.\ 2011; and references therein).

However, a proper comparison to observational datasets requires to convolve
model predictions with observational errors. We assume typical 1~$\sigma$ 
uncertainties of 0.3~mag in host luminosity, and 0.4~dex in BH mass, determined
via the virial technique
(Vestergaard \& Peterson 2006; Shen \& Kelly 2010; Bennert \etal 2011). 
The corresponding quantities in the SAM galaxy catalogue are
altered with randomly assigned errors in gaussian/lognormal distribution. 
The effects of error convolution are crucial, as shown in the bottom
panels of Fig.~\ref{fig:MBH_MRhost_obs}. The models now recover the 
observational results, although the most massive BH masses fall somewhat
short of the observed ones at the highest redshifts.

We find that it is the error on BH masses, rather than on host luminosities,
that has the main impact
in altering SAM predictions. This effect was discussed by Shen \& Kelly
(2010; see also Shen \etal 2008; Kelly \etal 2009): observational errors 
on measured BH masses, combined with the steep end of the BH mass function,
introduce a Malmquist--type bias that skewes the sample toward much larger 
apparent BH masses. We shall refer to this as the Shen--Kelly bias.
An analogous Malmquist--type bias at the bright end of the galaxy luminosity
function has proved to help to account for the stellar mass 
function of high $z$ galaxies in the hierarchical scenario
(Fontanot \etal 2009, and references therein).
 
The evolution of the BH mass--luminosity relation, in terms of brightening 
with redshift at given BH mass, is illustrated in Fig.~\ref{fig:DMR_evol}. 
The left panel shows the ``real'' evolution in the SAM: the global galaxy 
population gets steadily brighter at increasing redshift, and quasar hosts 
are predicted to be much brighter at any redshift.
The overluminosity depends on the BH mass:
around $M_{BH} \simeq 10^9$~\Msol\ --- the most interesting BH mass range 
for comparison with the dataset of Decarli \etal (2010ab) ---  the offset
is 0.7--1~mag, increasing at lower BH masses to almost 2~mag 
around $M_{BH} \simeq 10^8$~\Msol.

In the right panel we show the results after error convolution: while 
the evolution of objects around $10^8$~\Msol\ is marginally affected, 
the scenario drastically changes at the high mass end: for (apparent) BH masses
of $10^9 - 10^{9.5}$~\Msol, SAM are consistent with no evolution within
the errors, and become compatible with observational results.

Altogether, the combined effect of Lauer bias and Shen--Kelly bias
allow SAM to compare successfully to the observational results. 
Notice that both biases, acting at the massive/luminous end, produce
a steepening in the slope of the BH mass --- host luminosity (or host mass)
relation: the apparent slope is about 1.5~dex/mag.
Future observational investigations of the apparent slope, 
extending to QSOs of lower BH mass, will be a useful test for the models.

\begin{figure*}
\includegraphics[width=5.8truecm]{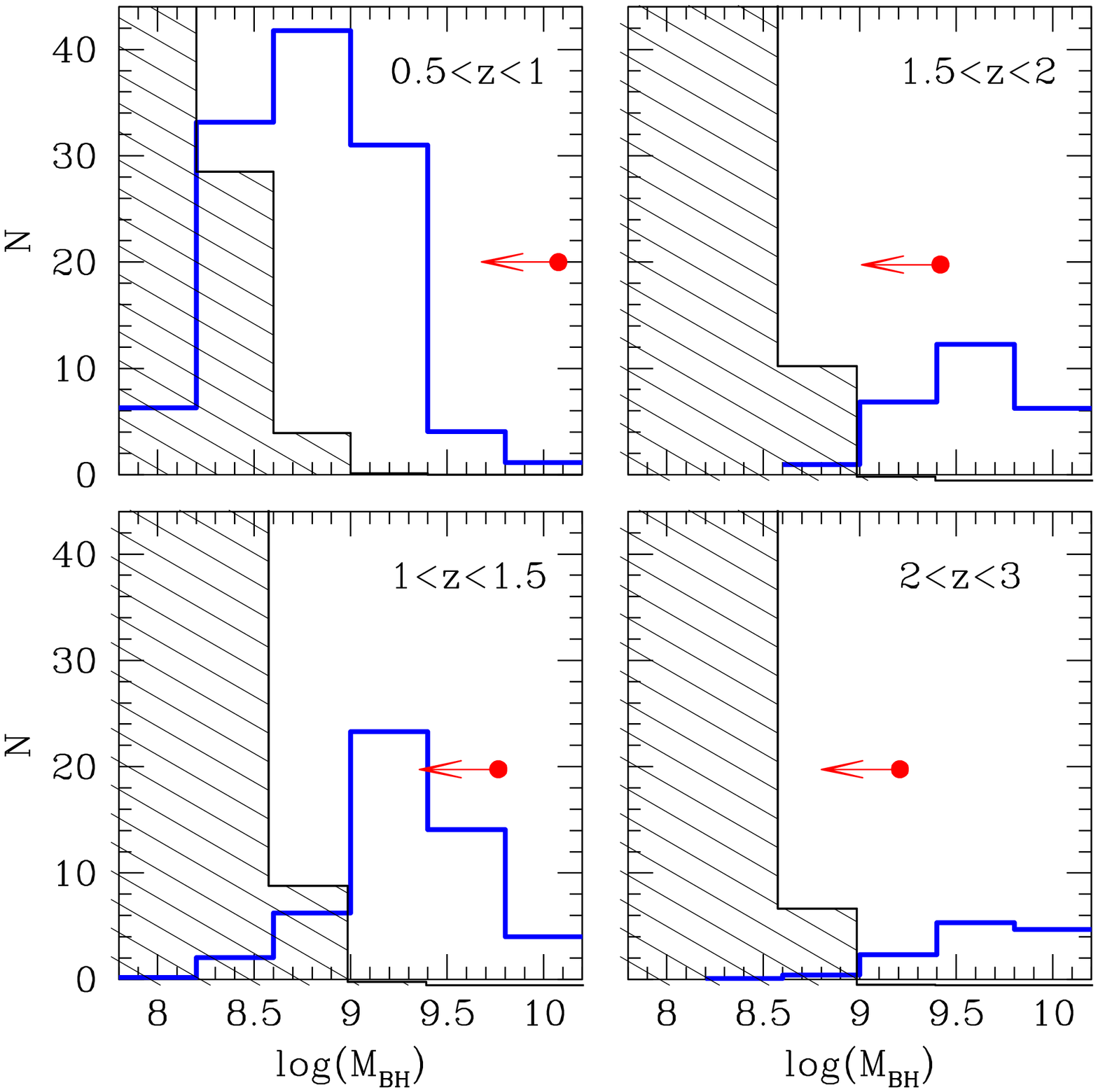}
\includegraphics[width=5.8truecm]{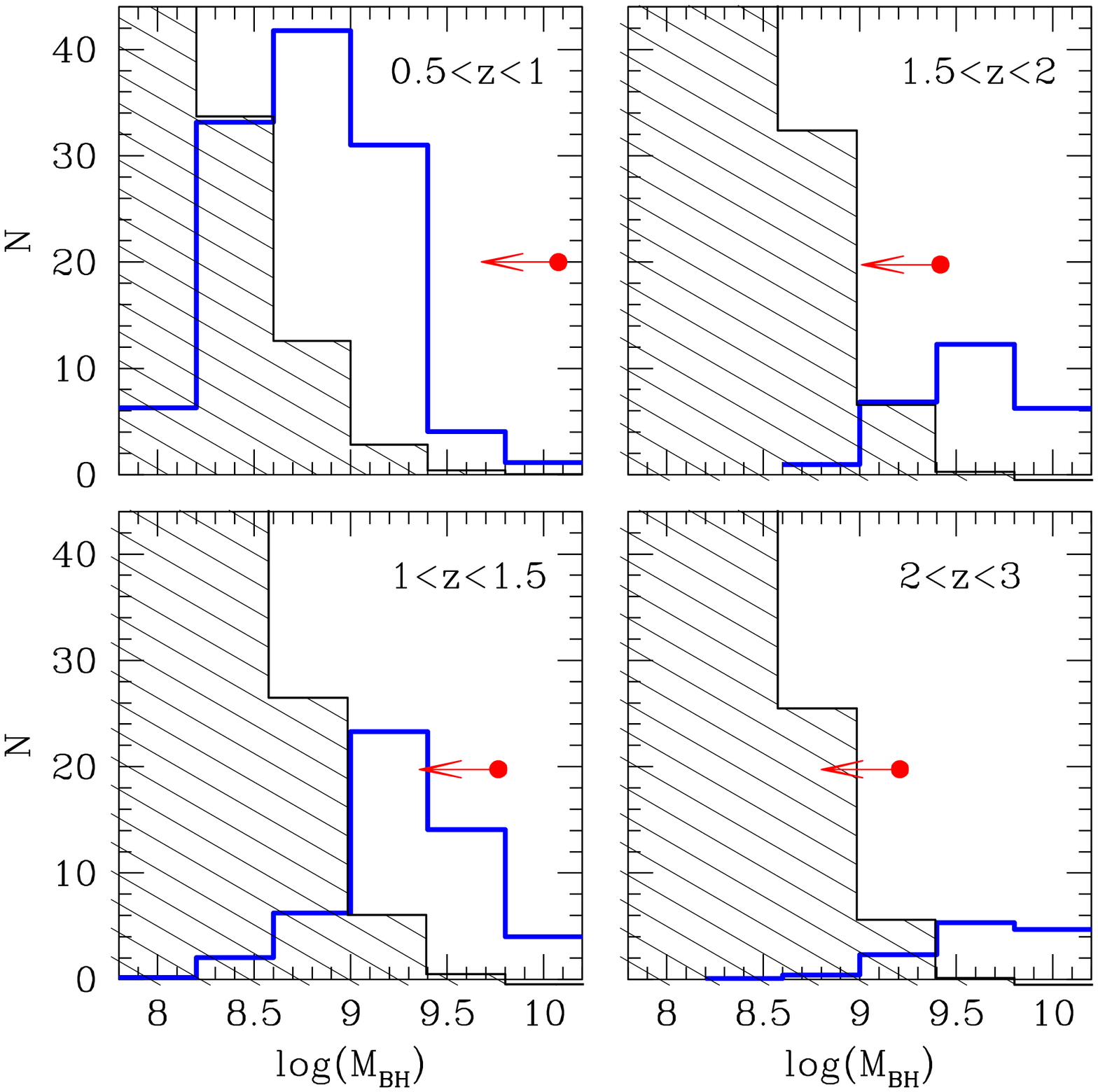}
\includegraphics[width=5.8truecm]{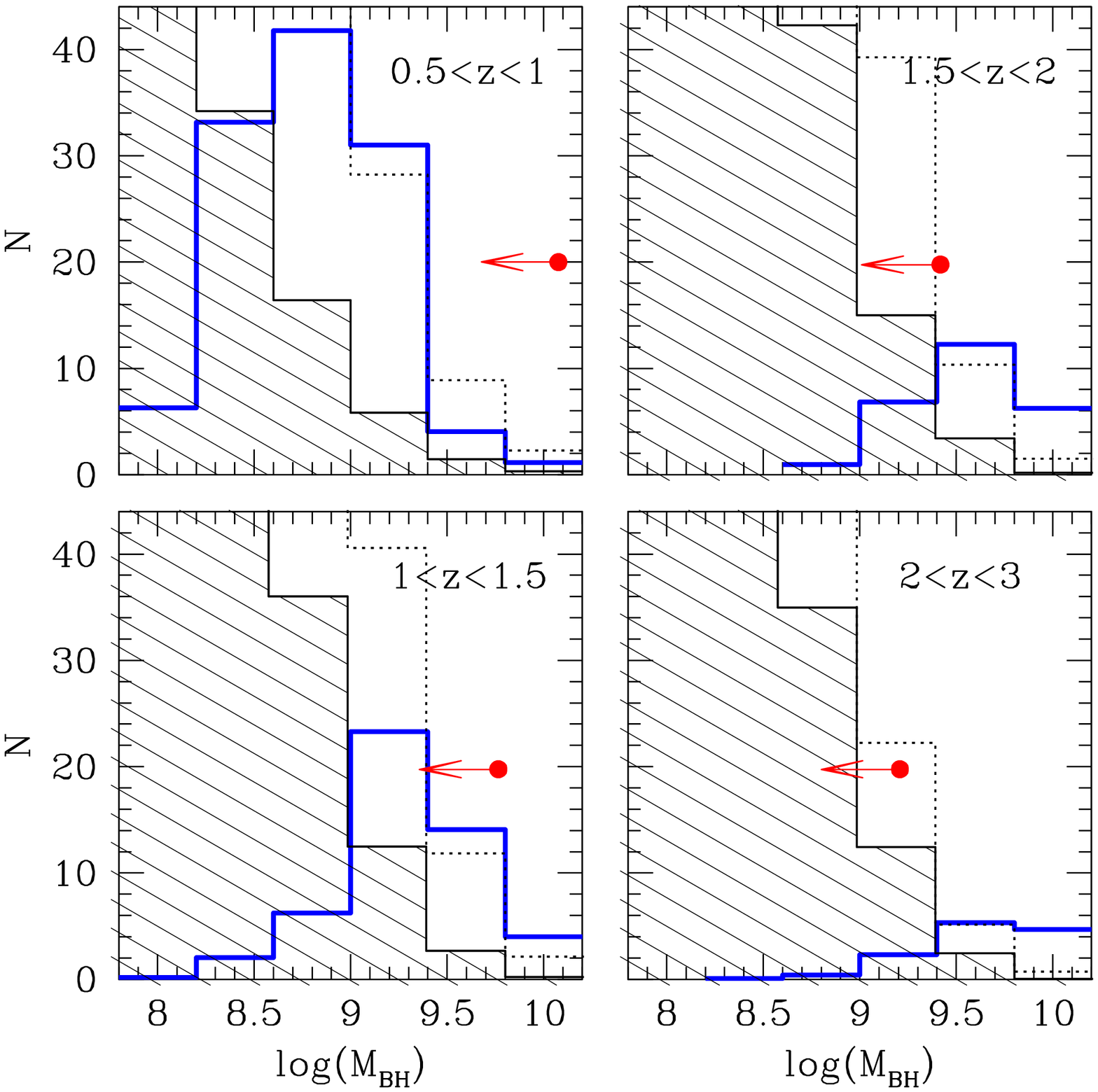}
\caption{{\it Thick (blue) histogram} : BH mass function of QSOs 
in a volume of the real Universe equal to that of the Millennium simulation 
(from Vestergaard \& Osmer 2009). 
{\it Thin shaded histogram} : BH mass function of selected QSO hosts, scaled 
considering that each of the selected merger galaxies in the redshift range 
indicated (corresponding to 1--2 Gyr of timespan) is active as an optical QSO 
for only $10^7$ yr. 
{\it Red dot with arrow}: maximum BH mass in the global galaxy population 
at the lowest end of the redshift bin (i.e.\ at $z=0.5, 1, 1.5$ and 2 for 
the various panels, respectively); it represents the maximum mass limit 
for QSOs that could possibly be active in that redshift bin; notice
the dearth of massive black holes ($M_{BH} \geq 10^9$) at $z>1.5$.
{\bf Left panel}: actual BH mass function in the SAM. {\bf Mid panel}: BH
masses of QSO hosts have been convolved with a lognormal error of 0.4~dex.
{\bf Right panel}: assuming a lognormal error of 0.55~dex; the {\it dotted 
histogram} is the (error--convolved) BH mass function of all merger galaxies 
(i.e.\ relaxing the ``doubling'' criterion).}
\label{fig:vester}
\end{figure*}

\section{The mass function of QSO's}
\label{sect:QSOmassfunc}

In the previous sections, we have seen how
statistical biases dominate the interpretation of the observed evolution
of the BH mass---host relation. 
Even with the ``aid'' of bias effects, though, Fig.~\ref{fig:MBH_MRhost_obs}
suggests that the SAM hardly reach the most massive BH observed in the
high redshift samples. Since the extent of the biases
strongly depends on the luminosity/mass function of galaxies and BH at the 
high mass end, we discuss in this section the observational constraints
on the mass function of QSO's. In particular, we consider whether the lack
of massive BH in the SAM is just a statistical limit, simply due 
to the fact that very massive quasars are too rare objects to be included 
in the simulation volume.

The Millennium simulation follows a comoving box of size 
500~$h^{-1}$=685~Mpc; from the mass function of quasars 
(Vestergaard \& Osmer 2009), in such a volume we expect about 10 active nuclei
with $10^{9.5} < M_{BH}<10^{10}$~\Msol\ at redshift $2<z<3$, while none is 
obtained in the simulations --- not only considering the selected quasar 
host galaxies, but even in the global galaxy population.
The left panel in Fig.~\ref{fig:vester} shows the number of expected 
active nuclei as a function of mass and redshift (thick histogram), 
compared to those obtained in the SAM. 
The excess of low mass QSOs in the SAM might depend on the details
of our selection criterion, or to the incompleteness of the observed QSO 
mass function below $10^9$~\Msol\ (Kelly \etal 2010). More important
for us here is the clear lack of quasars more massive
than $10^9$~\Msol\ at high redshift, independent of our selection 
criteria --- as it is confirmed looking at the global galaxy population.

This dearth of massive black holes at high $z$ may be due to an intrinsic
difficulty of hierarchical models to form massive objects at high redshift, 
or may demand a specific recipe for the formation of the most massive, rare BH.
Marulli \etal (2008) noticed an analogous mismatch with the bright end of the
AGN luminosity function at $z > 1$, and suggest that an accretion
efficiency increasing with redshift may cure the problem (see also Bonoli 
\etal 2009).
It remains to be seen how the new prescription would impact the evolution
of the scaling relations and the Lauer and Shen--Kelly bias in the Munich SAM.
Both biases are strongest at the high mass/luminosity end, therefore
a BH mass function depleted already at $M_{BH}=10^9$~\Msol\ probably corresponds
to an enhanced bias at that BH mass.

However, here also we must convolve model predictions with 
realistic observational errors. In the middle panel of Fig.~\ref{fig:vester} 
we show the results after convolving model BH masses with a lognormal
error distribution of 0.4~dex standard deviation, similar to that adopted
in Section~\ref{sect:hostlum}. The comparison with the observed mass function
at the high mass end is improved, yet not satisfactory: the problem
of undermassive BH persists, at least above $M_{BH}=10^{9.5}$~\Msol. 
(Notice that no error convolution is considered on the dot--with--arrow, 
i.e.\ on the most massive BH
actually formed in the simulation; this highlights how the Shen--Kelly bias
on QSO hosts can produce even higher BH masses, than actually existing
in the whole simulated volume.)

However, if typical errors as large as 0.55~dex are allowed for the virial
technique (Vestergaard \& Osmer 2009; Vestergaard 2010; but see also Kelly 
\etal 2010, favouring smaller uncertainties), the discrepancy
between SAM and observed mass function is much reduced (right panel in
Fig.~\ref{fig:vester}). Especially relaxing the ``doubling'' criterion 
on BH masses for
the selection of QSO hosts (see Section~\ref{sect:selection}; dotted histogram)
and taking into account that cosmic variance is typically 2--3 times Poisson 
noise.
All things considered, there is some evidence for a lack of massive BH in
simulated QSO hosts, but it is not compelling once observational errors
are included. A deeper investigation on this issue would require a detailed
simulation of the QSO light curves and luminosities, so as to extract from the
SAM a sample of objects mimicking closely the observational selection.

Finally we remark that, while the Shen--Kelly bias depends only on the BH
mass function and the uncertainties on measured BH masses, the Lauer bias 
is also sensitive to the luminosity function of galaxies: a paucity of 
simulated luminous, massive galaxies at high redshift would also enhance 
this bias. In this respect,
we notice that the long--standing difficulty of most SAM with 
the K--band galaxy luminosity function at early epochs, seems to be now 
overcome thanks to improved treatment of the critical AGB phase in population 
synthesis models (Henriques \etal 2011).

\section{Discussion and conclusions}

QSO host galaxies at high redshift are important tracers of the
co--evolution of galaxies and black holes. 
Taking advantage of recent datasets extending out to $z=3$, we have studied 
how the observed evolution of the BH --- host scaling relations compares 
to theoretical semi--analytical models; we considered specifically 
the publicly available SAM of the Munich group (De Lucia \& Blaizot 2007).

While at $z=0$ the scaling relations are established for the general galaxy
population, at high $z$ BH masses can be only derived for active
nuclei by means of the virial technique. This introduces a number of
potential biases, to be taken into account when discussing the 
evolution of the scaling relations.
\begin{enumerate}
\item
Quasar host galaxies are in a peculiar phase of their evolution: in the
theoretical scenario considered here, they are ``young spheroids''
that have just merged and suffered a starburst. Our analysis
highlights the distinction between the general population and the recent 
mergers/quasar hosts.
\item
At high redshift it is hard to decompose the host galaxy into its bulge/disc 
component so the scaling relations we analyze refer to the global galaxy;
yet, for consistency with observational papers, evolution is defined with
respect to the local relations derived for quiescent host spheroids.
\item
Luminous quasars tend to trace over--massive BH with respect 
to the underlying intrinsic BH--host relation (Lauer \etal 2007), so
the comparison relation in the models must be defined accordingly.
\item
The observational errors on BH masses introduce a Malmquist--type bias
(Shen \& Kelly 2010) that also must be taken into account, by convolving
model prediction with observational errors before direct comparison
to the data.
\end{enumerate}
We find that the latter two bias effects dominate the interpretation of
the observational results. In the Munich SAM, two basic predictions are:
(i) the intrinsic (bisector--fit) relation between BH mass and host stellar 
mass has negligible evolution out to $z=3$
--- as typical of models that do not include quasar feedback and 
self--regulation mechanisms; (ii) quasar host galaxies are systematically 
overluminous
(and/or have systematically undermassive black holes) with respect
to the local BH mass --- host luminosity relation. Both predictions, taken
at face value, are in stark contrast with observations. 
However, the Lauer bias in the SAM produces an apparent evolution 
of 0.6~dex out to $z=3$, for the host stellar mass of black holes with
$M_{BH} \sim 10^9$~\Msol (the typical BH masses probed by high redshift QSOs):
this is comparable to the observed evolution of $\Gamma$ 
(Section~\ref{sect:hostmass}). Besides, when observations and models are
directly compared in the 
BH mass---host luminosity plane, and models are properly convolved with
observational errors, the Shen--Kelly bias compensates for the intrinsic
overluminosity of SAM quasar hosts, bringing the models into agreement with 
the observations (Section~\ref{sect:hostlum}).

We thus find that the observed strong evolution, with BH
formation preceding the growth of the hosts, could be largely the 
result of statistical and selection biases, compatible with negligible real
evolution of the intrinsic BH mass -- host mass relation; this agrees with 
the conclusion of Shen \& Kelly (2010).
Whether a strong $\Gamma$ evolution really characterizes the general 
co--evolution of BH and galaxies, is therefore still unclear.
We note, for instance, that sub-mm galaxies tend to trace the opposite trend 
($\Gamma$ decreasing at high $z$), which can be understood if different 
selection biases apply to different sub--populations of galaxies
(Lamastra \etal 2010, and references therein).

Since biases dominate the interpretation of the results, it is  
of paramount importance to ascertain that SAM predict realistic biases. 
As both the Lauer and the Shen--Kelly bias are related to the fact that
high--$z$ quasars trace the massive/bright end of the BH and galaxy
distribution functions, SAM should reproduce these adequately 
at various redshifts.
While the situation for the galaxy luminosity function is nowadays satisfactory
(Henriques \etal 2011), there is evidence that the Munich SAM
fail at reproducing the high mass end of the BH mass function at early epochs.
Indications for this come from the bright end of the
AGN luminosity function at $z > 1$ (Marulli \etal 2008) and from the mass
function of high--$z$ QSOs (Section~\ref{sect:QSOmassfunc}); though this latter
evidence is less compelling, if an error on observed BH masses 
as large as 0.55~dex is allowed and cosmic scatter is considered. A deeper
investigation on this issue requires more detailed modelling of the BH accretion
history and QSO luminosity curves, so as to extract from the SAM catalogue 
QSO samples that closely mimic the observational datasets.

A dearth of massive black holes ($M_{BH}>10^{9.5}$~\Msol) in the simulated volume
may be due to a general difficulty of hierarchical galaxy formation models 
to produce massive objects at high redshift, or to the fact that these 
massive black holes are so rare (e.g.\ Decarli \etal 2010b) that 
a separate, specific scenario is required to implement their formation
in SAM. Alternative mechanisms of BH formation in the very high redshift
Universe, advocated to account for the rarest, most massive quasars at 
$z \simeq 6$
(e.g.\ Mayer \etal 2010, and references therein) may indeed help also to 
improve on the statistics of massive quasars at $z$=3 and below.

Progress in the interpretation of high redshift data also requires a better
understanding of the biases in the real Universe. Both the
Lauer bias and the Shen--Kelly bias act at the high end of the BH mass function,
producing a steepening of the apparent BH mass --- host relation with respect
to the intrinsic one. Both effects are predicted to vanish around
$M_{BH} \leq 10^8$~\Msol, and to be present also at low redshifts. Therefore,
assuming evolution to be negligible at relatively low redshifts, comparing 
the relation for the local galaxy population to that for AGN hosts, 
can constraint the actual biases.
Also extending high--redshift samples to lower BH masses would be
valuable.

In summary, the interpretation of the properties of quasar hosts involves 
a full account of the statistical 
properties (luminosity/mass functions) of both galaxies and quasars:
on one hand quasar hosts are useful tests for SAM, on the other hand
we need the global approach of SAM to properly interpret the data.
The SAM considered here, although not adequately reproducing 
the AGN population, can still recover the observed trend of $\Gamma(z)$ 
in quasar host galaxies, when selection biases are included; and suggests
that the underlying $\Gamma$ evolution for the general galaxy 
population, may be much milder. It will be worthwhile to reconsider the role 
of biases at the massive end of the BH populations, in the context 
of SAM that better account for the properties of the quasar population.

The available observational datasets at present consist of a relatively 
small number of objects, but larger samples are expected to become available 
in the near future, based on high resolution observations with the next 
generation of 30--50~mt.\ telescopes. 
We conclude with a ``wish--list'' for future semi--analytical studies,
to fully exploit the potential of quasar hosts galaxy observations 
to constrain the co--evolution of BH and galaxies.

\begin{itemize}
\item
SAM should include the modelling of the quasar accretion rate and light curve, 
so as to predict the properties of galaxies and the BH--host relations 
{\it specifically during the phase of optical quasar activity}, as in
Kauffmann \& Haehnelt (2000).\footnote{Detailed BH accretion histories 
and AGN light curves 
have been modelled within the Munich SAM by Marulli \etal (2008); their 
effect is minor on the final scaling relations, where the total accreted 
BH mass matters more than the accretion 
timescale. The accretion history, though, affects the properties
of the host versus the instant observed quasar luminosity. This type of result
was discussed by Kauffmann \& Haehnelt (2000, their Fig.~12 and~18) but,
to our knowledge, by no other more recent SAM paper, from any research group.}
\item
In analyzing the co--evolution of the BH mass and its host, a clear distinction
should be made between intrinsic (bi--sector fit) relation and median relation 
at a given BH mass. The latter is affected by the Lauer bias, whose 
effects should be assessed separately. Error convolution, including the
Shen--Kelly bias, is another mandatory step.
\item
Besides the $M_{BH}-M_\star$ relation, SAM should provide predictions
on the $M_{BH}-L$ relation, which allows a more fair and self--consistent 
comparison to the observations.
\end{itemize}

Effort is particularly required to reproduce properly the mass/luminosity 
function of quasars at high redshift at the massive end: due to the 
importance of statistical biases, this is a crucial pre-requisite 
to our understanding of the co--evolution of BH and galaxies as traced 
by quasar hosts.

\section*{Acknowledgments}
We thank Pierluigi Monaco for constructive comments and suggestions; 
Gabriella De Lucia and Gerard Lemson for clarifying various aspects
of the SAM galaxy catalogue in the Millennium database; Peter Johansson
and Marianne Vestergaard for useful  dicussions. 
This study was financed by the Academy of Finland (grants nr.~219317 and 
2600021611) and by the Italian Ministry for University and Research (MIUR).
The Millennium Simulation databases used in this paper and the web application 
providing online access to them were constructed as part of the activities 
of the German Astrophysical Virtual Observatory.

\section*{Appendix : Colour and mass--to--light ratio evolution of quasar hosts}
\label{sect:colour}

The observed luminosities of quasar host galaxies are to be translated
into stellar mass, in order to recover the underlying BH mass --- host mass
relation to be compared to the local one. In this Appendix we discuss
the stellar mass--to--light ratio ($M_\star/L$) necessary for the transformation.

A passively evolving starburst formed at $z=5$ well describes the observed 
dimming of quasar hosts (Kotilainen \etal 2009) and was consequently assumed 
by Decarli \etal (2010b) to convert luminosities to stellar masses. Similar
assumptions were made by Peng \etal (2006). Let us compare 
the colour and $M_\star/L$ evolution predicted by the SAM, to the classic 
assumption of passive evolution.

SAM galaxies are expected to be bluer and have lower
$M_\star/L$ than a passively evolving galaxy,
since in a hierarchical Universe galaxies build up progressively and are
on average younger than in the monolithic scenario. Quasar hosts,
selected to be recently merged objects with associated starbursts, should
deviate even further from passive evolution. 

\begin{figure}
\centering
\includegraphics[width=7.7truecm]{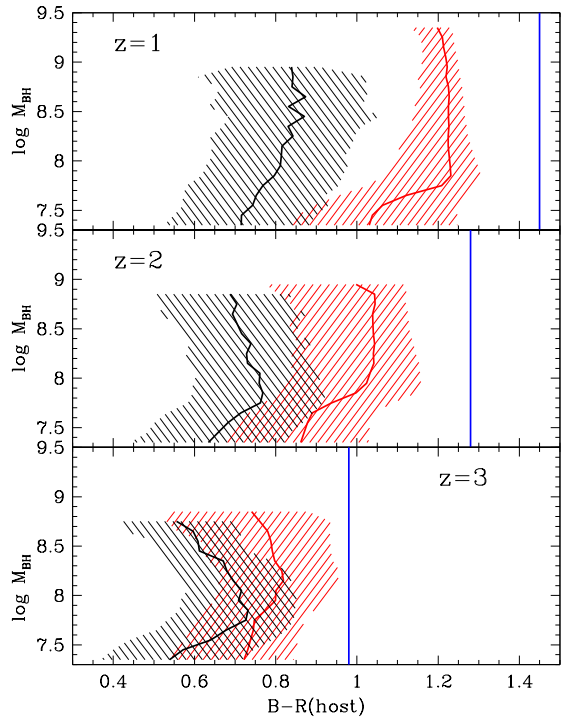}
\caption{Colour distribution of galaxies at three redshift snapshots. 
The leftmost solid line, with shadings inclined to the left,
represent the median and the 16 and 84 percentiles for the quasar host
galaxies. The (red) solid line in the middle, with shadings inclined 
to the right, represents the analogous for the global population. 
The (blue) vertical line to the right shows the colours of a passively 
evolving starburst formed at $z=5$.}
\label{fig:B-Rcolour}
\end{figure}

Fig.~\ref{fig:B-Rcolour} shows the $(B-R)$ colour distribution of SAM galaxies 
as a function of redshift. Both for the quasar hosts and for the global galaxy 
population, the typical colours are quite independent of the central BH mass 
above {\mbox{$M_{BH} \geq 10^8$~\Msol}} (i.e.\ the median lines are roughly 
vertical in the plot). At $z \leq 1$ there is a significant offset 
in colour between the global average galaxy population and the quasar hosts,
that are systematically bluer by about 0.4 mag due to 
merger--induced recent star formation. At increasing redshift the offset  
decreases, as the global population gets on average bluer, faster than the 
quasar hosts; by $z=3$, the offset is reduced to $<$0.2~mag, corresponding 
to only $1 \sigma$ difference between the two populations. The vertical (blue)
line shows, for comparison, the much redder colours expected for passive 
evolution since $z=5$.

\begin{figure}
\includegraphics[angle=-90, width=8.8truecm]{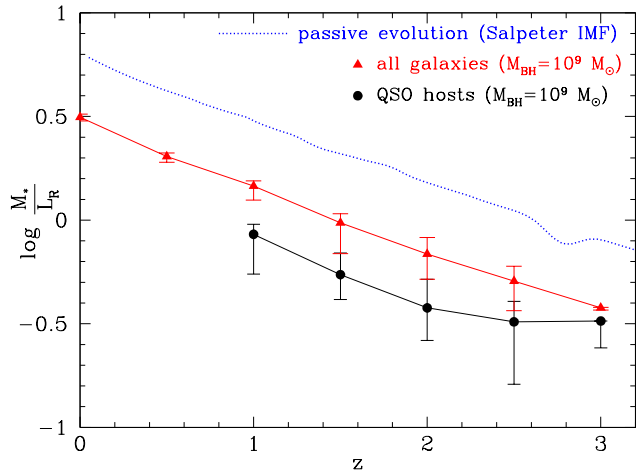}
\caption{Evolution of the stellar mass--to--light ratio in rest--frame $R$ band
for SAM galaxies with a central BH mass of $10^9$~\Msol (all galaxies and QSO
hosts, respectively). The dotted line is the passively evolving $M_\star/L$
adopted by Decarli \etal (2010b) to transform observed luminosities into stellar
masses.}
\label{fig:MLRevol}
\end{figure}

Fig.~\ref{fig:MLRevol} shows the evolution of the $M_\star/L$ ratio
in rest--frame $R$--band, for the QSO hosts and the global galaxy 
population respectively. We also draw the mass--to--light of a passively 
evolving starburst formed at $z=5$, computed by Decarli \etal (2010b) with 
the aid of the GALAXEV package of Bruzual \& Charlot (2003)  
to convert their observed luminosities to stellar masses. 
Interestingly, the rate of $M_\star/L_R$ evolution of the SAM galaxies 
is very similar to the passively evolving scenario; the offset of 0.3~dex
can be partly ascribed to the different stellar IMF adopted 
(Salpeter 1955 for passive evolution, Chabrier 2003 for the SAM galaxies); and
partly to the fact that SAM galaxies are significantly bluer than a purely
passively evolving galaxy (Fig.~\ref{fig:B-Rcolour}). Quasar hosts also define 
an evolutionary rate mimicking passive evolution, at least up to $z<2.5$, 
with a further offset of 0.2~dex.

As the {\it rate} of luminosity evolution is similar in the various scenarios, 
the result of Decarli \etal (2010b) that quasar hosts were significantly
undermassive at high redshift does not strongly depend on the passive evolution
assumption. Actually, adopting the lighter $M_\star/L$ ratios predicted by the
SAM would only strengthen their findings, with central BH being even more
overmassive, by a further 0.3--0.5~dex, with respect to their hosts.

The behaviour shown in Fig.~\ref{fig:MLRevol} also highlights that a complex 
galaxy formation history may easily mimic a passively evolving case when 
viewed in a monochromatic band.\footnote{Another example of this is found in 
the evolution of the K--band luminosity function (Cirasuolo \etal 2007, 2010): 
the characteristic luminosity of the Schechter function, $M_{K,\star}(z)$, 
brightens with redshift following the passive evolution of a high--redshift 
starburst, so as to apprently trace a population of ellipticals formed 
at $z>3$. However, when the authors consider the decrease in number density 
of bright galaxies beyond $z=1.5$, and the evolution of the red and blue 
populations separately, the apparent passive fading of $M_{K,\star}$ clearly
hides a much more complex galaxy evolution history.}
A possible way to distinguish a truly passively
evolving population from a merger scenario is to use colour information
(Fig.~\ref{fig:B-Rcolour}). Unfortunately, multi--band information on quasar 
hosts at high $z$ is still scarse and mostly limited to $z \lsim 1.5$ 
(Jahnke \etal 2009; Bennert \etal 2011). 
Moreover, since the host luminosity and colors have typical uncertainty of 
0.3~mag one can hardly discriminate between the two scenarios 
beyond $z \sim 2$.

As to the adopted IMF for the $M_\star/L$ normalization, most recent
theoretical models of galaxy formation adopt the ``bottom--light'' 
Chabrier (2003) prescription; however, for the most massive ellipticals 
that presently host the most massive BH --- analogous to those traced 
by high redshift QSO --- recent results 
suggest that a Salpeter, or even ``heavier'' IMF, may be more appropriate
(Treu \etal 2010; Thomas \etal 2011; Van Dokkum \& Conroy 2010, 2011; 
Tiret \etal 2011). 
The direct comparison in the BH mass--host luminosity plane 
(Section~\ref{sect:hostlum}), however, bypasses the transformation problem.


\end{document}